\documentclass[10pt,prx,aps,twocolumn,floatfix]{revtex4}
\usepackage{graphicx}
\usepackage{amsmath}
\usepackage{epstopdf}
\usepackage{braket} 
\usepackage{bm} 
\usepackage{color}
\usepackage{amsfonts}

\begin{document}

\title{Tuning the topological insulator states of artificial graphene}

\author{H. D. Scammell and O. P. Sushkov}
\affiliation{School of Physics, The University of New South Wales, Sydney, New South Wales 2052, Australia}
\date{\today}

\begin{abstract}
We develop a robust, non-perturbative approach to study the band structure of {\it artificial graphene}. Artificial graphene, as considered here, is generated by imposing a superlattice structure on top of a two dimensional hole gas in a semiconductor heterostructure, where the hole gas naturally possesses large spin-orbit coupling. Via tuning of the system parameters we demonstrate how best to exploit the spin-orbit coupling to generate 
time reversal symmetry-protected topological insulator phases. Our major conclusion is the identification of a second set of topological Dirac bands in the band structure (with spin Chern number $C=3$), which were not reliably obtainable in previous perturbative approaches to artificial graphene. Importantly, the second Dirac bands host more desirable features than the previously studied first set of Dirac bands (with $C=1$). Moreover, we find that upon tuning of the system parameters, we can drive the system to the highly desirable regime of the {\it topological flat band}. We discuss the possibilities this opens up for exotic, strongly correlated phases. 
\end{abstract}

\maketitle

\section{Introduction}
The quantum spin Hall effect, exhibited by time reversal symmetric topological insulators (TI), is by now a well investigated topic. It has the desirable property of dissipationless spin currents along the edges of the sample, which is a fundamental ingredient for reversible quantum computation as well as for efficient spintronics applications. Realisations of two-dimensional (2D) TI states has been slow, with the best known examples arising in HgTe \cite{Konig2007} and InAs \cite{Knez2011} quantum wells as predicted theoretically in Refs. \cite{BernevigScience2006, Bernevig2006, Zhang2006}. As is well known, graphene was the first proposed material realisation of a TI \cite{Kane2005}, however within naturally occurring graphene the spin-orbit interaction is parametrically small. Therefore the topological gap, a measure of the stability of the TI state against (time reversal symmetric) disorder and thermal fluctuations, is small. 

Aside from generating a robust TI state for the sake of technological advances, there has been a recent surge of interest in 2D topological band insulators for the purpose of realising highly non-trivial strongly correlated phases of matter. In particular, the theoretical efforts have been focused on designing/predicting systems exhibiting (nearly) flat bands with non-trivial topology. Such conditions are expected to be sufficient to realise novel correlated phases: fractional Chern- \cite{Tang2011, Neupert2011,  Hu2011, Bernevig2011, Kourtis2012, Regnault2012}, fractional anomalous- \cite{Neupert2011,Wang2011, Venderbos2012} and fractional topological- insulators \cite{Stern2009, NeuportPRB2011, Sun2011, Sheng2011, Zhong2013} (see reviews \cite{Goerbig2012, PARAMESWARAN2013}), magnetic insulating phases \cite{Katsura2010, Santos2012, Hohenadler2013, Kumar2014, Doretto2015} (review \cite{Hohenadler2013}), or superconducting/superfluid phases \cite{Peotta2015, Kauppila2016, Tovmasyan2016, Iskin2017, Liang2017}. The logic is rather simple, the flat band implies that kinetic energy (which vanishes) is dominated by particle-particle interactions, even if they are `weak'. Moreover, the flat bands support a macroscopic degeneracy -- large density of states. Partial filling of such a flat band therefore becomes an exciting playground for strongly-correlated physics. The prototypical example is the fractional quantum Hall effect, where the flat bands are the exactly flat Landau levels. Very recently, by analogy with the fractional quantum Hall effect, there are mounting theoretical efforts to explain/predict fractional Chern Insulators, fractional anomalous Hall effect, and fractional TIs. To date, the pursuit for nearly flat bands with non-trivial topology has led to the proposal of several model Hamiltonians with at times peculiar properties such as complex or long range hopping parameters, for which finding an experimental realisation is a formidable task and requires fine tuning.

The present work considers a graphene simulator, {\it artificial graphene}, comprised of a two dimensional hole gas (2DHG) confined by a semiconductor quantum well, with an electrostatic potential of hexagonal symmetry etched onto a metallic top-gate. This {\it superlattice} structure generates a graphene-like electronic band structure. There have been numerous graphene simulator proposals; cold atoms \cite{Sols2008, Soltan-Panahi2011, Tarruell2012, Uehlinger2013}, lithographic \cite{Singha2011, Park2009, Gibertini2009, Potemski2012, Ghosh2012, Pellegrini2010}, and more \cite{Haldane2008, Bittner2012, Rechtsman2013, Gomes2012}, for a review see \cite{Polini2013}. 
Our primary motivation is to optimise (with respect to system parameters) the robustness of the topological insulator phase, i.e. the topological gap. 
To this end, a hole gas, as opposed to an electron gas, is the obvious choice since the holes in semiconductor heterostructure systems posses effective spin $3/2$ angular momentum, and thus naturally experience larger spin-orbit coupling than the corresponding spin $1/2$ electrons. Our secondary motivation is to search for nearly flat bands with non-trivial topology. Due to the nature of the dispersion of the 2DHG we will see that generating a nearly flat band once the superlattice is imposed becomes a natural feature of the artificial graphene spectrum -- and does not require fine tuning. 

Artificial graphene (as defined here) is a readily tunable system and hence has already been proposed as a candidate material to exhibit: a TI state \cite{SushkovNeto2013}, a Chern insulator under (in-plane) applied field \cite{Li2016}, as well as realising a topological semimetal \cite{Li2017}. We are concerned with the TI state suggested in \cite{SushkovNeto2013}, whereby such calculations have been based on a purterbative theory of the 2DHG -- valid in the limit  of small spin-orbit interaction -- which then enters into the band structure calculations of the superlattice, artificial graphene. Although these perturbative approaches no doubt capture the essential qualitative physics, in the limit of large spin-orbit, which is the most desirable for technological application or to pursue the topological flat band regime, they become unreliable. 
To address this is gap in the literature, the present work develops a non-perturbative description of artificial graphene, valid at arbitrarily large spin-orbit coupling. Hence, as experimental efforts move closer to explicit realisation, the necessity of the present work is self-evident. 

The next section, Section \ref{Prelims}, provides a self-contained introduction to the technical aspects of the paper. In which we outline: the construction of artificial graphene; the necessary technical details of the of the 2DHG; the previous perturbative approach \cite{SushkovNeto2013}; and finally we develop our new nonperturbative approach to artificial graphene. Having established these necessary preliminaries we move onto our results in Sections \ref{PT} and \ref{P Broken}. 


\begin{figure}[t!]
 {\includegraphics[width=0.435\textwidth,clip]{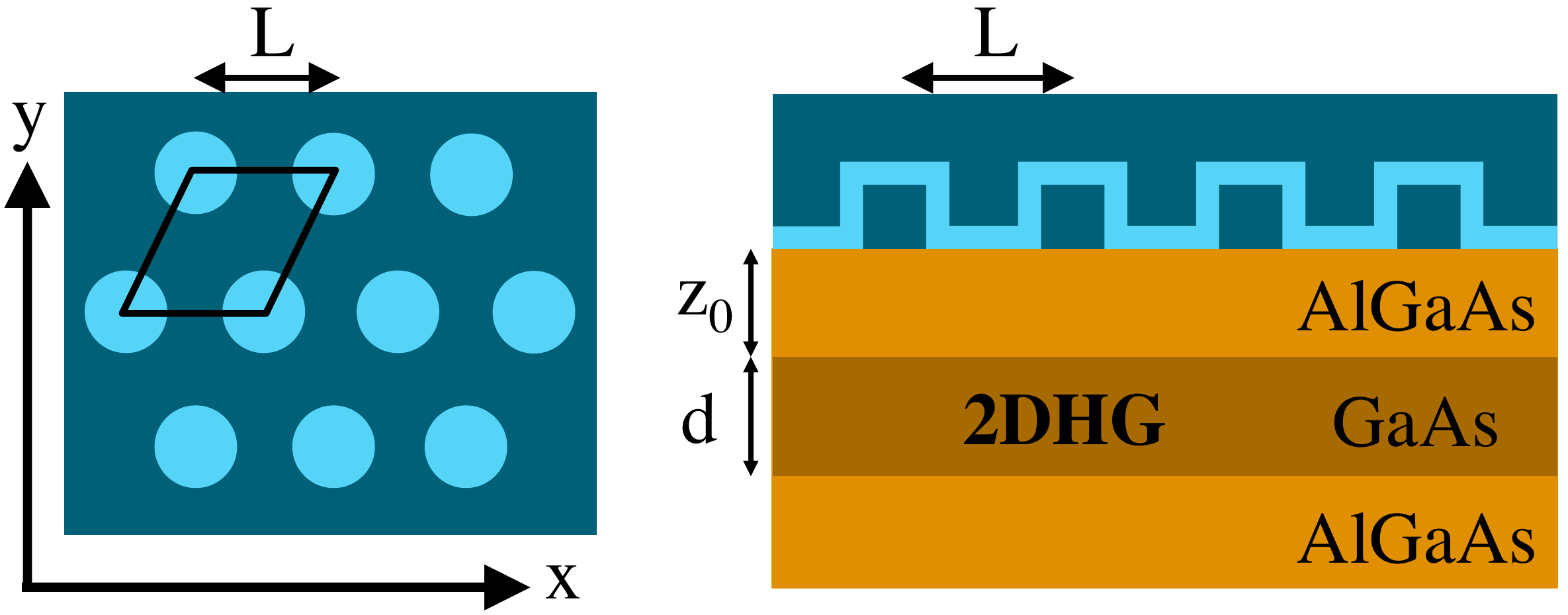} \vspace{0cm}}
 \caption{Schematic view of artificial graphene. (a) A top view of the superlattice etched into the the metallic top-gate. Blue dots represent positive potentials (anti-dots). $L$ is the lattice spacing. (b) A cross section of the AlGaAs-GaAs-AlGaAs heterostructure. $d$ is the quantum well confined length, $z_0$ is the separation length between the superlattice top-gate and the 2DHG. }
 \begin{picture}(0,0) 
\put(-9,90){\text{(b)}} 
\put(-115,90){\text{(a)}} 
\end{picture}
\label{Schematics}
\end{figure}
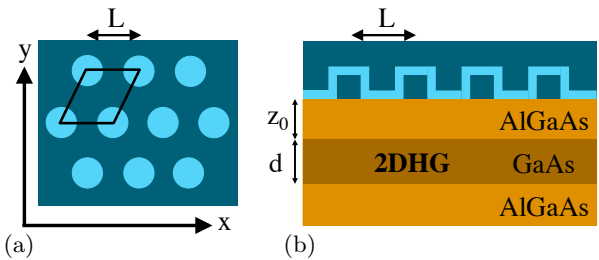

\section{Mathematical Preliminaries}\label{Prelims}
\subsection{Key parameters of artificial graphene}\label{params}
To adequately set the stage for the mathematical details to follow, we briefly outline the schematics of artificial graphene, and in doing so establish the key parameters available for tuning, see Figure \ref{Schematics}. First, we consider a 2DHG having in mind e.g. AlGaAs-GaAs-AlGaAs quantum well. The confinement is along the z-axis, leaving the x-y plane for free motion; this is the 2DHG. For the bulk of the results/calculations we take the confinement to be a rectangular quantum well of width $d$. We will also perform semi-analytics for the case of a triangular well. Next, a periodic electrostatic potential of triangular symmetry is etched onto a metal plate on top of the 2DHG; this is the superlattice. The lattice parameter (henceforth the {\it superlattice parameter}) is $L$. The separation along the z-axis of the superlattice top-gate from the 2DHG is $z_0$. Although $z_0$ plays a role \cite{Tkachenko2015}, we will fix its value and not consider it further. Finally, we denote the magnitude of the electrostatic potential by $W$. Ultimately, it is the ratio of $d/L$ that controls the spin-orbit interaction, and we can choose $W$ freely; this provides us with two tuning handles: $d/L$ and $W$. Also note that tuning $z_0$ is equivalent to tuning $W$.

\subsection{2DHG}\label{2DHG}
We consider a 2DHG confined along the $z$-axis by the quantum well potential
\begin{gather}
\label{sqconfine}
V(z) =  
\begin{cases}
  0, & z\in(-d/2,d/2)\\    
 \infty, &\text{otherwise.}    
\end{cases}
\end{gather}
For this confinement we set the characteristic momentum and energy scale to be
\begin{align}
\label{sqaurescales}
k_0&=\frac{2}{d}, \ \ \ E_0=\frac{\gamma_1 k_0^2}{2m}\equiv\frac{k_0^2}{2m^*},
\end{align}
where $m$ is the electron mass in vacuum, $\gamma_1$ is one of the Luttinger parameters entering the Hamiltonian below, and we have introduced the effective mass $m^*$ to facilitate later discussion. The energy scale for a quantum well of width $d =20$ nm is $E_0 = 2.6$ meV. It is important to note that this energy scales as $E_0\sim1/d^2$.

The holes posses an ultra-relativistic spin-orbit coupling, and can be described by the Luttinger Hamiltonian in the axial approximation, i.e. $U(1)$ symmetry in-plane. The axial approximation is useful for quasi-2D systems with frozen dynamics along
one direction, in the present case, the z-axis. The Luttinger Hamiltonian we consider is \cite{Miserev2017},
\begin{align}
\label{HL}
H_\text{2DHG}&=H_0+H_\text{SO},\\
\notag H_0&=\left(\gamma_1+2\gamma_2\left(\frac{5}{2}- S_z\right)\right)\frac{k_z^2}{2m}\\
\notag &+\left(\gamma_1-\gamma_2\left(\frac{5}{4}- S_z\right)\right)\frac{\bm k^2}{2m} + V(z),\\
\notag H_\text{SO}&=-\frac{\gamma_2+\gamma_3}{8m}\left(k_+^2S_-^2 + k_-^2S_+^2\right)\\
\notag&- \frac{\gamma_3}{4m}\{k_z,\{S_z,k_+S_- + k_-S_+\}\},
\end{align}
where $S_x, S_y, S_z$ are angular momentum 3/2 operators, $S_{\pm}=S_x \pm i S_y$ and we use bold font to express the in-plane momenta $\bm k =(k_x,k_y)$, and $k_{\pm}=k_x \pm i k_y$. In the expression for $H_\text{2DHG}$ (\ref{HL}), we have chosen to separate the components $H_0$ and $H_\text{SO}$ (which accounts for spin-orbit interactions). This is merely to help illustrate the following technical step: we perform exact diagonalization of $H_\text{2DHG}$ in the basis of wavefunctions obtained from $H_0$. A later step will be to project the wave functions of $H_\text{2DHG}$ onto the superlattice potential -- the details will be provided in section \ref{Superlattice theory}.

Some effort has been made \cite{Miserev2017} to obtain semi-analytic expressions for the $H_\text{2DHG}$ wavefunctions, which saves computational time and provides a clearer mathematical picture. We outline the results obtained previously \cite{Miserev2017}: the wave functions are labeled by a $\bm k$ (a good quantum number) as well as the energy-level index $l$ and corresponding spin index $\sigma_l$, which corresponds to the physical spin projection only at $\bm k=0$. The wavefunctions of $H_\text{2DHG}$ for quantum well confinement in $z$-direction read
\begin{align}
\label{ket}
\ket{l, \sigma_l, \bm k}&=e^{i\bm{k r}}\sum_{S_z}\hat{k}_+^{(\sigma_l-S_z)}\sum_n a_{l,n,S_z}(k)\ket{S_z,n},
\end{align}
where, importantly, the argument of $a_{l,n}$ is $k\equiv|\bm k|$, and all phase dependence on the 2D momentum plane is stored in the pre-factor $\hat{k}_+^{(\sigma_l-S_z)}$ such that $\hat{k}_{\pm}\equiv (k_x \pm i k_y) /|\bm k|$, which greatly simplifies the band structure computations. The ket, $\ket{S_z,n}$, accounts for the basis of the four $S_z=\pm 3/2,\pm 1/2$ spinors and the infinite set of harmonics, enumerated by the non-negative integers $n$, due to confinement along the $z$-axis.

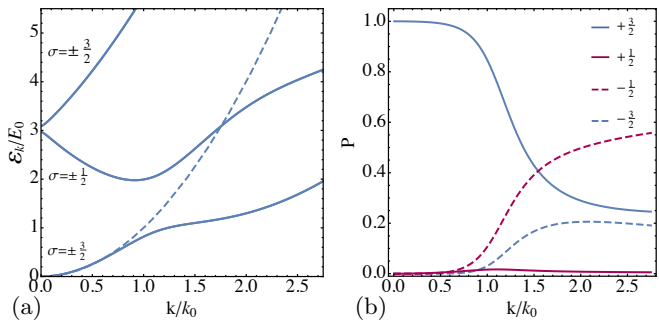
\begin{figure}[t!]
 {\includegraphics[width=0.235\textwidth,clip]{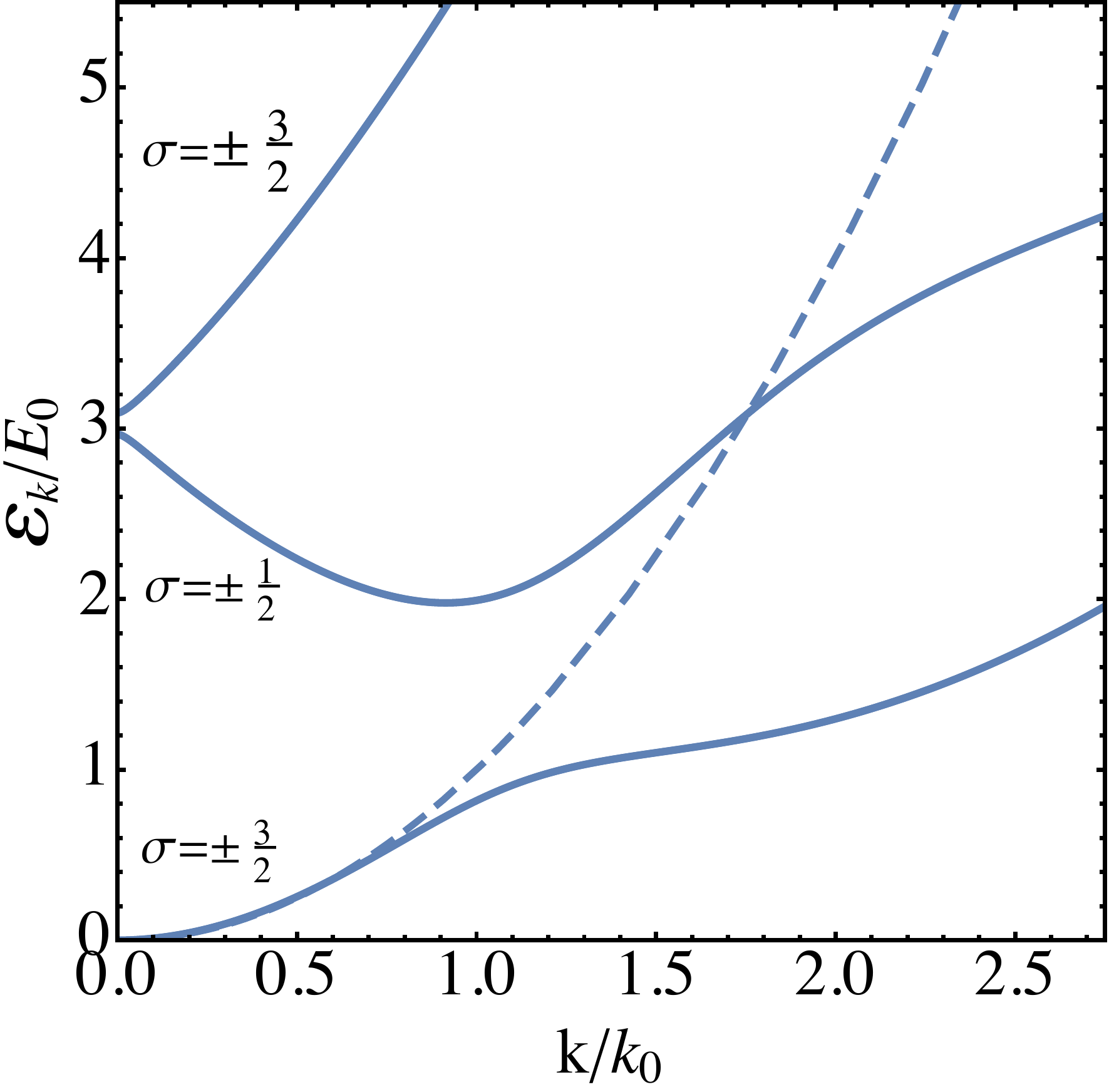}} \hspace{0cm}
{\includegraphics[width=0.235\textwidth,clip]{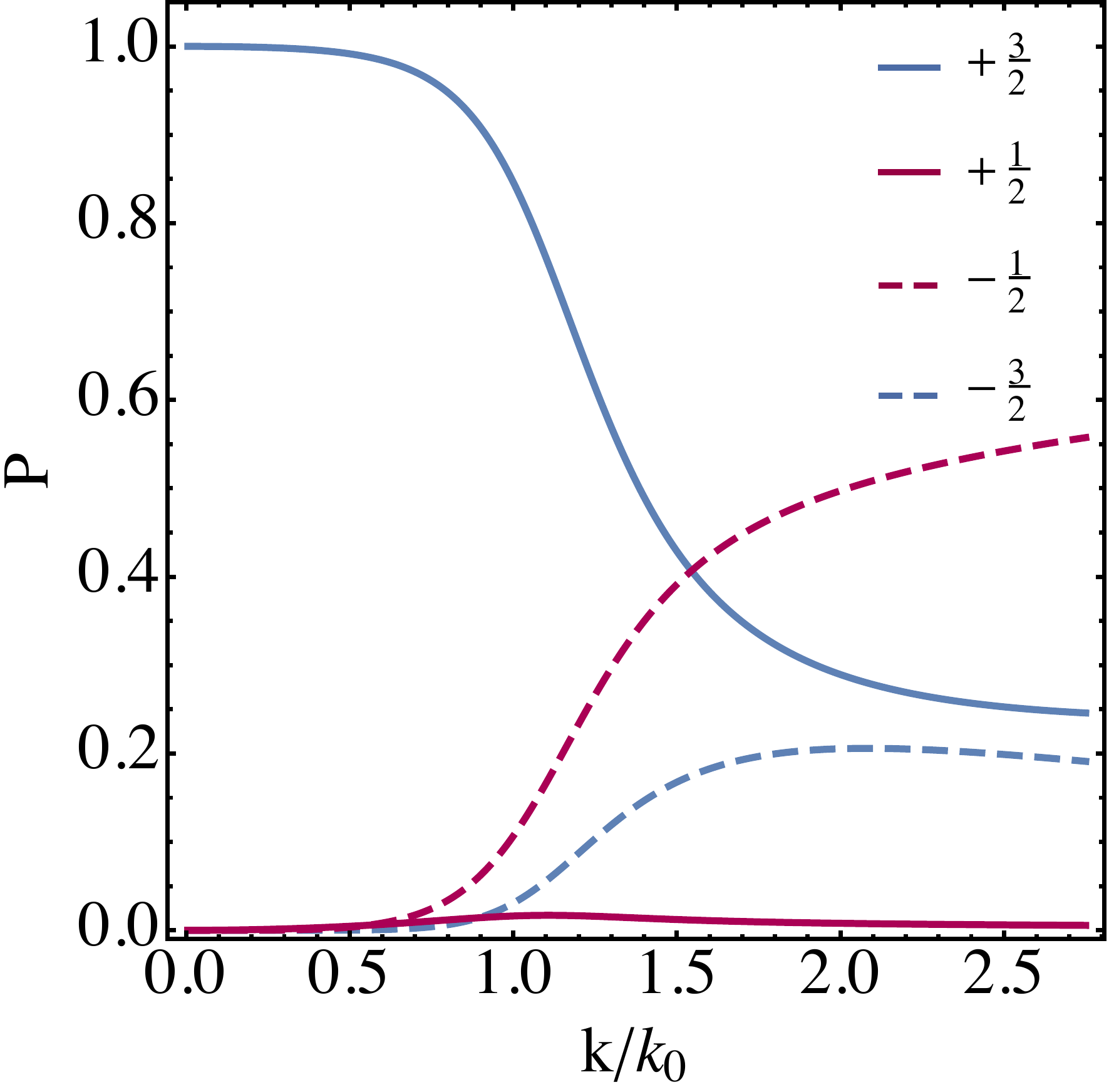}} \hspace{0cm}
 \caption{(a) The dispersion of holes in the semiconductor quantum well:  Solid lines are the dispersions obtained from the multi-band Luttinger Hamiltonian approach (\ref{HL}). The dashed line corresponds to single-band, quadratic approximation, i.e. using ${\cal E}_k=k^2/(2m^*)$. The projections of the total angular momentum are denoted $\sigma$, which at $k=0$ are identically the physical spin projections. (b) A plot of the relative weight of the projection onto each physical spin component $S_z$, and their dependence on $\bm k$, in the lowest band Kramers spin state $\ket{l=0,\sigma_l=+3/2,\bm k}\equiv\ket{\uparrow,\bm k}$.}
\begin{picture}(0,0) 
\put(10,140){\text{(b)}} 
\put(-120,140){\text{(a)}} 
\end{picture}
\label{2DHGspectrum}
\end{figure}

To be explicit, Figure \ref{2DHGspectrum}(a) shows the first three dispersion levels $l=0,1,2$ and the corresponding spin indices $\sigma_l=\pm 3/2, \pm 1/2, \pm 3/2$ (for the quantum well potential). If one considers just the $l=0$ sub-space, then it is enlightening to regard this as a pseudo-spin 1/2, such that $\ket{l=0, +3/2, \bm k}=\ket{\uparrow, \bm k}$ and $\ket{l=0, -3/2, \bm k}=\ket{\downarrow, \bm k}$. Due to time reversal symmetry (denoted ${\cal T}$): ${\cal E}_{\uparrow, \bm k}={\cal E}_{\downarrow, \bm -k}$, i.e. $\ket{\uparrow, \bm k}$ and $\ket{\downarrow, \bm k}$ are Kramers partners. (Note: throughout the text we call the two possible angular momentum eigenstates/projections of any given dispersion band $l$ the {\it Kramers spin}.) Moreover, for the quantum well confinement there is an inversion symmetry or parity (${\cal P}$), such that:  ${\cal E}_{\uparrow, \bm k}={\cal E}_{\uparrow, \bm -k}$. 

Figure \ref{2DHGspectrum}(b) shows the probabilities of each physical spin ($S_z$) component of the particular state $\ket{l=0, +3/2, \bm k}=\ket{\uparrow, \bm k}$. The probabilities of each $S_z$ are expressed as (using notation introduced in Eq. (\ref{ket}))
\begin{align}
\label{probabilities}
P_{l, \sigma_l,S_z}&=\sum_n |a_{l,n,S_z}(k)|^2,
\end{align}
which are normalized for each index $\{l, \sigma_l\}$ such that $\sum_{S_z}P_{l, \sigma_l,S_z}=1$. We see from Figure \ref{2DHGspectrum}(b) that all spin components are mixed via $\hat{J}=1,2,3$ (dipole, quadrupole, octupole).

It is also instructive to contrast the exact diagonlization results of Eq. (\ref{ket}) (and also partially represented in Figure \ref{2DHGspectrum}(b)), to results of perturbation theory \cite{SushkovNeto2013}.  From perturbation theory, the wavefunctions of the $l=0$ Kramers doublet are given as,
\begin{align}
\label{PertWaveFuncs}
\ket{\uparrow, \bm k}&=\left[\ket{+\frac{3}{2}} + \beta \bm k_+^2\ket{-\frac{1}{2}}\right]e^{i\bm k\cdot\bm r}, \\
\notag\ket{\downarrow, \bm k}&=\left[\ket{-\frac{3}{2}} + \beta \bm k_-^2\ket{+\frac{1}{2}}\right]e^{i\bm k\cdot\bm r}, \\
\notag\beta&=\frac{\sqrt{3}d^2}{4\pi}.
\end{align}
Hence, in this approach we clearly see that only mixing via $\hat{J}=2$ (quadrupole) selection occurs, i.e. $S_z=\pm 3/2$ and $\mp 1/2$. Moreover, the wavefunctions (\ref{PertWaveFuncs}) are valid only if $\beta^2 k^4\ll1$.

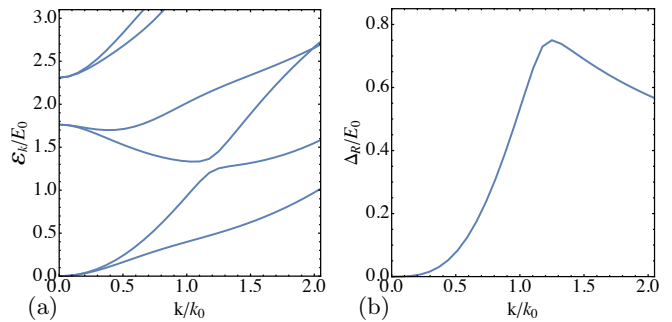
\begin{figure}[t!]
 {\includegraphics[width=0.235\textwidth,clip]{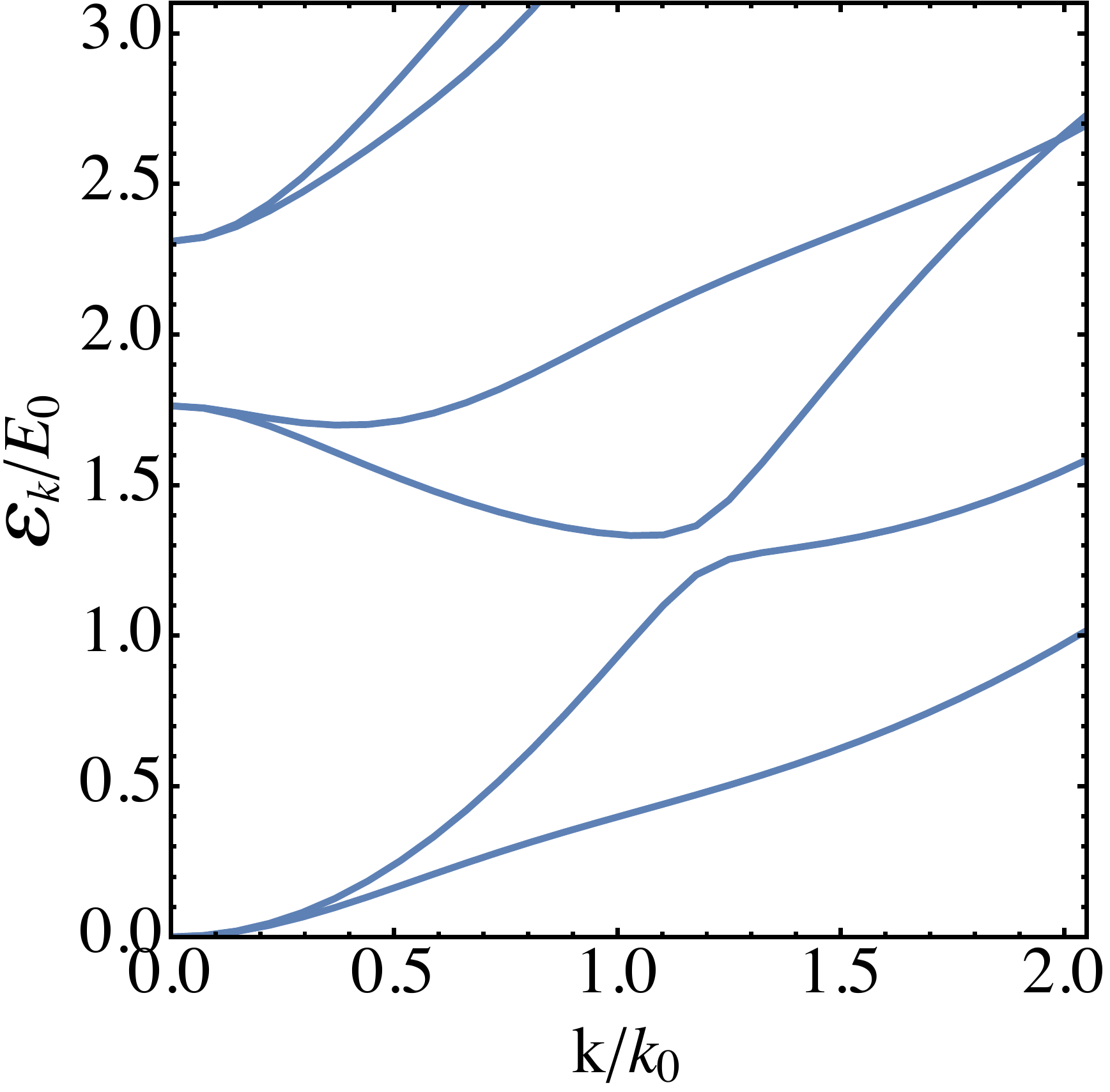}} \hspace{0cm}
  {\includegraphics[width=0.235\textwidth,clip]{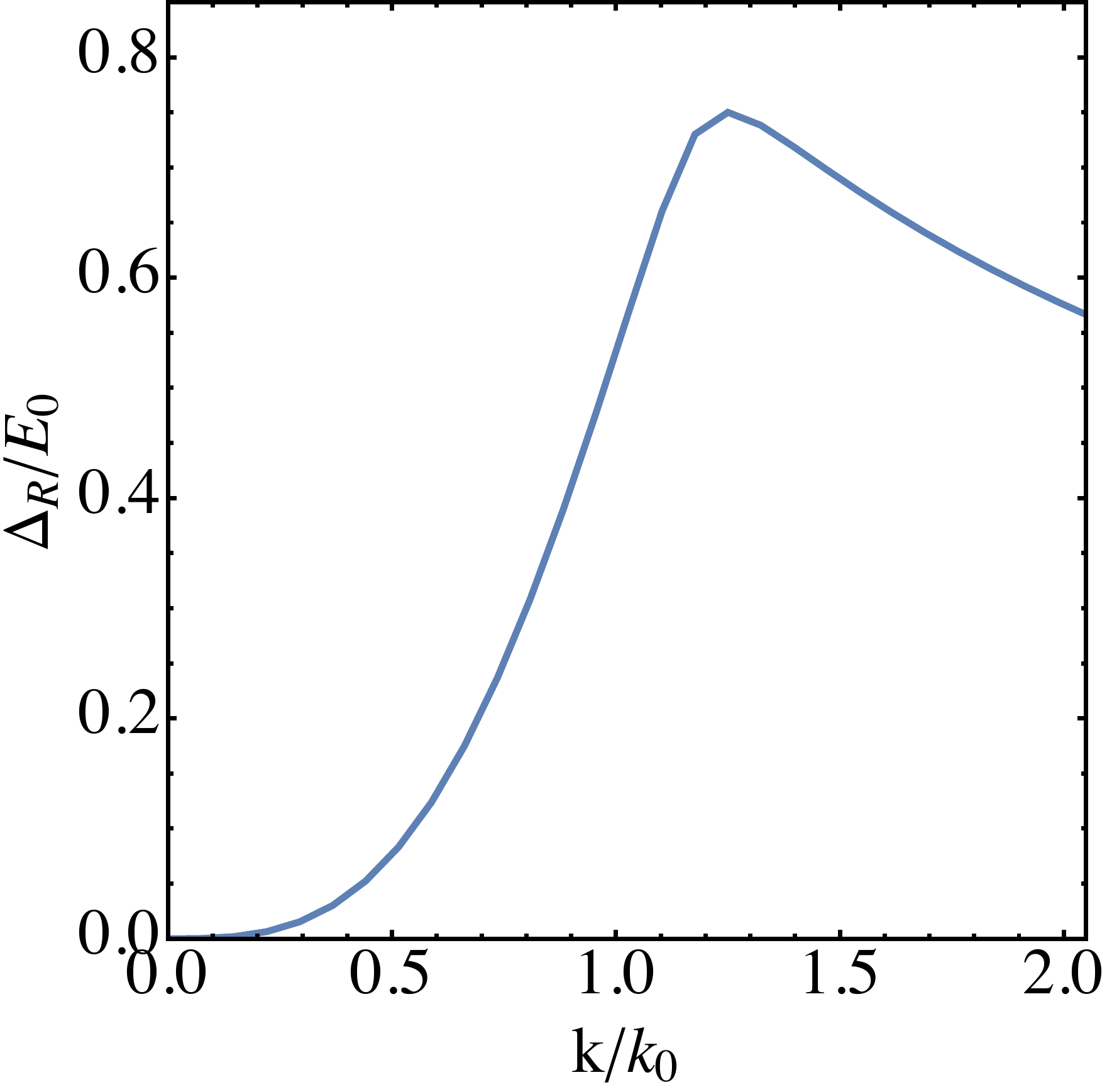}} \hspace{0cm}
 \caption{(a) Spectrum of 2DHG in the triangular well confinement of Eq. (\ref{triconfine}). (b) Rashba splitting: the energy splitting between the two lowest ($l=0$) dispersion bands in (a).}
 \begin{picture}(0,0) 
\put(-115,67){\text{(a)}} 
\put(10,67){\text{(b)}} 
\end{picture}
\label{Triangularspectrum}
\end{figure}

We will also consider a triangular well confinement by introducing the confinement potential 
\begin{gather}
\label{triconfine}
V(z) =  
\begin{cases}
  e{\cal E} z, & z>0\\    
 \infty, &\text{otherwise.}    
\end{cases}
\end{gather}
For this confinement we set the characteristic momentum and energy scale to be
\begin{align}
\label{triscales}
k_0&=\frac{1}{2}me{\cal E}, \ \ \ E_0=\frac{\gamma_1 k_0^2}{2m}.
\end{align}
This potential explicitly breaks ${\cal P}$ and with it, ${\cal E}^{\text{tri}}_{\uparrow, \bm k}\neq{\cal E}^{\text{tri}}_{\uparrow, \bm -k}$, which removes the double degeneracy of the energy at a given momentum. We perform exact diagonalization of the Luttinger Hamiltonian Eq. (\ref{HL}) subject to the triangular confinement; the energy spectrum is shown in Figure \ref{Triangularspectrum}(a). In Figure \ref{Triangularspectrum}(b) we show the energy splitting between the two $l=0$ bands, we call it the {\it Rashba splitting} and denote it by $\Delta_R$. The important point here is that ${\cal P}$-breaking confinement Eq. (\ref{triconfine}) generates the Rashba spin-orbit interaction. In section \ref{P Broken} we will take a semi-analytic approach and consider just these two lowest bands in an effective Hamiltonian.



\begin{figure}[t!]
{\includegraphics[width=0.235\textwidth,clip]{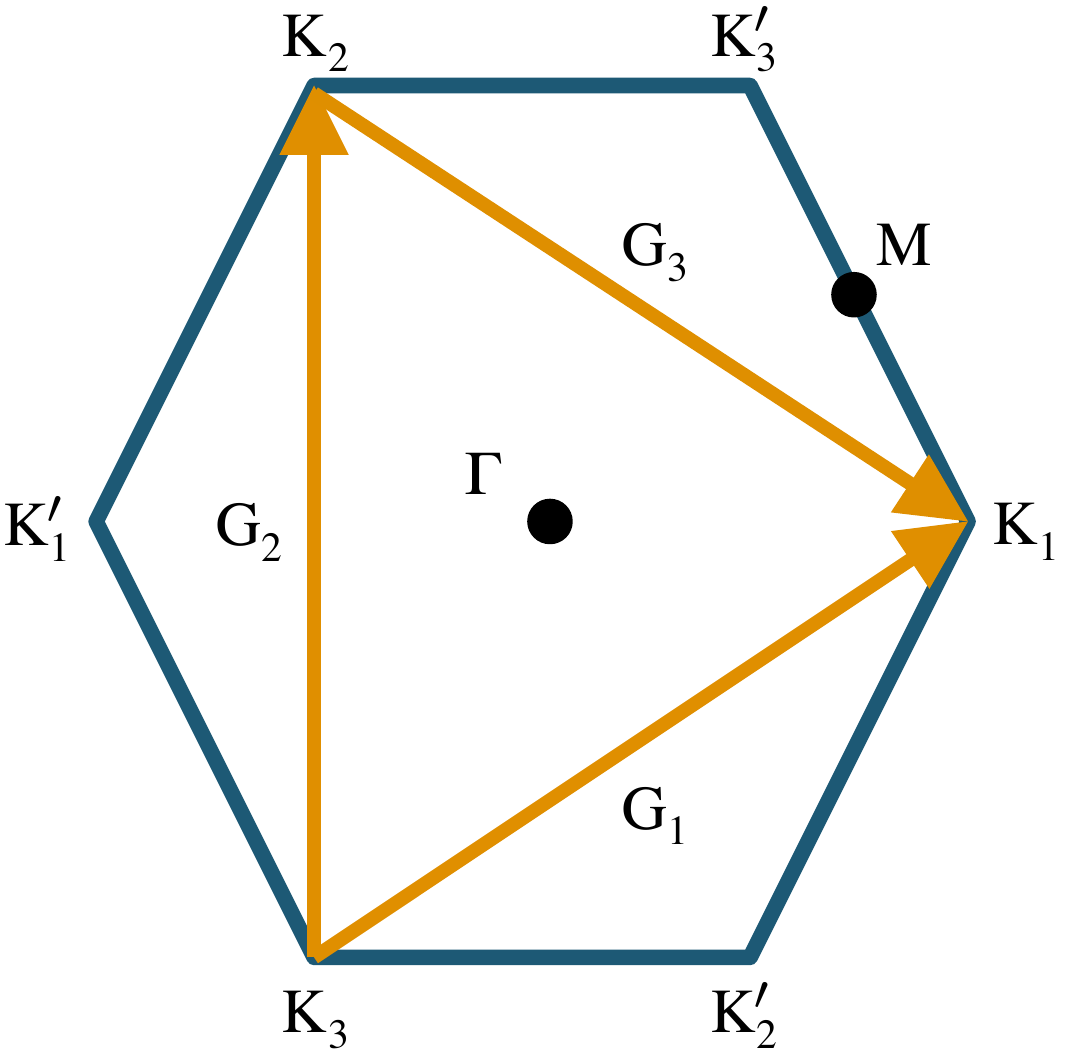}} \hspace{0cm}
 \caption{Brillouin zone. $\bm G_i$ are vectors connecting zone corners $\bm K_j$. $\bm K_j'$ are parity reflections of $\bm K_j$}
\label{BZ}
\end{figure}

\subsection{Superlattice theory}\label{Superlattice theory}
We describe the superlattice potential via a harmonic approximation (for more details see Ref. \cite{Tkachenko2015}),
\begin{align}
\label{U}
{\cal U}({\bm r})&=2W\sum_{t=1}^3 \cos({\bm G}_t\cdot{\bm r}),
\end{align}
where $W$ is a constant that determines the strength of the potential. The vectors ${\bm G}_i$ ($|{\bm G}_i|=4\pi/(\sqrt{3}L))$ are the reciprocal lattice vectors connecting corners of the hexagonal Brillouin zone ${\bm K}_j$ ($|{\bm K}_j|=4\pi/(3L)$), as shown in Figure \ref{BZ}. It is convenient to introduce the energy scale of the superlattice,
\begin{align}
\label{EL}
E_L&=\frac{{\bm K}_j^2}{2m^*}=\frac{8\pi^2}{9L^2m^*}.
\end{align}

We describe the problem by the {\it superlattice Hamiltonian} operator,
\begin{align}
\label{Hoperator}
\hat{{\cal H}}&=\hat{\cal E}_\text{2DHG} + \hat{{\cal U}}(\bm r),
\end{align}
whereby $\hat{\cal E}_\text{2DHG}$ represents the {\it kinetic energy} operator, which encodes the dispersions of holes in the 2DHG, i.e. its matrix elements (in a given basis) are the eigenvalues of $H_\text{2DHG}$ (\ref{HL}), also shown in Figure \ref{2DHGspectrum}(a). The potential $\hat{{\cal U}}$ is given by Eq. (\ref{U}), except we use a hat/operator notation to imply that we need to project these operators onto a particular basis. 
We take the wave functions of $H_\text{2DHG}$ (\ref{HL}) as this basis, and project the superlattice Hamiltonian $\hat{{\cal H}}$ onto them. Previously in Eq. (\ref{ket}) we used notation $\ket{l, \sigma_l, \bm k}$ for the $H_\text{2DHG}$ wavefunctions, now due to the superlattice potential $\hat{{\cal U}}(\bm r)$ we must attach an extra index $i$ (such that $\ket{l, \sigma_l, \bm k}\to\ket{l, \sigma_l, \bm k, i}$) that labels sites in the momentum grid $\bm k_i=\bm k + \bm g_i$, where the discrete momentum space grid $\bm g_i \in \{n_1 \bm G_1 + n_2 \bm G_2 + n_3 \bm G_3 : n_i\in Z\}$, is the space of degenerate momentum points. Note, $\bra{l, \sigma_l, \bm k, i}\hat{\cal E}_\text{2DHG}\ket{m, \sigma_m, \bm k, j}$ is diagonal in all indices and $\bra{l, \sigma_l, \bm k, i}\hat{\cal U}(\bm r)\ket{m, \sigma_m, \bm k, j}$ is traceless (since it is traceless in the indices $i,j$).

\subsubsection{Perturbative, single-band theory}\label{PertDirac}
Let us now apply the procedure outlined above to the simplest case: considering just three degenerate points (of the same parity) ${\bm K}_1,{\bm K}_2,{\bm K}_3$, we project the superlattice Hamiltonian (\ref{Hoperator}) onto the perturbative wavefunctions of Eq. (\ref{PertWaveFuncs}), which have just $l=0, \ \sigma_{l=0}=\pm3/2$ components, and find 
\begin{align}
\label{H33}
{\cal H}_{i,j,l,m}&=\bra{\sigma_l, \bm k,i}\hat{\cal H}\ket{\sigma_m, \bm k,j}\\
\notag({\cal H}_{i,j,l,m})& = 
  \begin{pmatrix}
     {\cal E}({\bm k} + {\bm K}_1) & W & W  \\
    W & {\cal E}({\bm k} + {\bm K}_2) & W \\
    W & W & {\cal E}({\bm k} + {\bm K}_3)
  \end{pmatrix} \otimes\mathbb{I}\\
\notag& + \frac{1}{\sqrt{3}} \eta
  \begin{pmatrix}
    0 & i & -i  \\
    -i & 0 & i \\
    i & -i & 0
  \end{pmatrix}\otimes\tau_z
\end{align}
where $\mathbb{I}$ is the two-dimensional identity matrix and $\tau_z$ is the usual Pauli matrix, both of which act on the physical spin/Kramers doublet subspace. In this approximation, the diagonal elements are just quadratic dispersions ${\cal E}({\bm k})={\bm k}^2/(2m^*)$ -- we refer to this as the {\it single-band, quadratic approximation}. The coefficient $\eta=3/2 \beta^2 K_1^4 W$, with $\beta$ taken from Eq. (\ref{PertWaveFuncs}) and $W$ from Eq. (\ref{U}), determines the strength of the spin-orbit coupling. Consider ${\bm k}=0$, whereby the diagonal elements are equal since $|{\bm K}_j|=4\pi/(3L)$, upon setting $\eta=0$ we find that there is a doubly degenerate eigenvalue $\{-W,-W,2W\}$ of ${\cal H}$ in (\ref{H33}): this is the Dirac point \cite{Park2009,SushkovNeto2013}. Projecting the Hamiltonian (\ref{H33}) onto the doubly degenerate subspace, allowing for $\eta\neq0$, and performing a small ${\bm k}$-expansion about ${\bm K}_1,{\bm K}_2,{\bm K}_3$ gives the Kane-Mele-like Hamiltonian \cite{SushkovNeto2013},
\begin{align}
\label{Dirac}
H_D&=-v(p_x\sigma_y +p_y\sigma_z)\otimes\mathbb{I}-\eta\sigma_x\otimes\tau_z,
\end{align}
where $\sigma_i$ are Pauli matrices acting on the psuedo-spin space generated by the doubly degenerate eigenvalues $\{-W,-W\}$ of ${\cal H}$ in (\ref{H33}).
From the Dirac-like Hamiltonian (\ref{Dirac}) one finds \cite{SushkovNeto2013} that the spin-orbit gap is given by $\Delta_{SO}=2\eta\sim(d/L)^4W$.
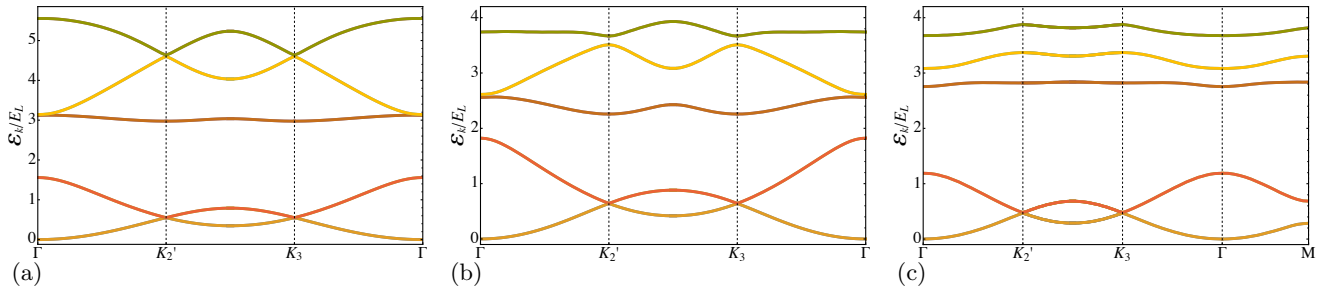
\begin{figure*}[t!]
 {\includegraphics[width=0.3223\textwidth,clip]{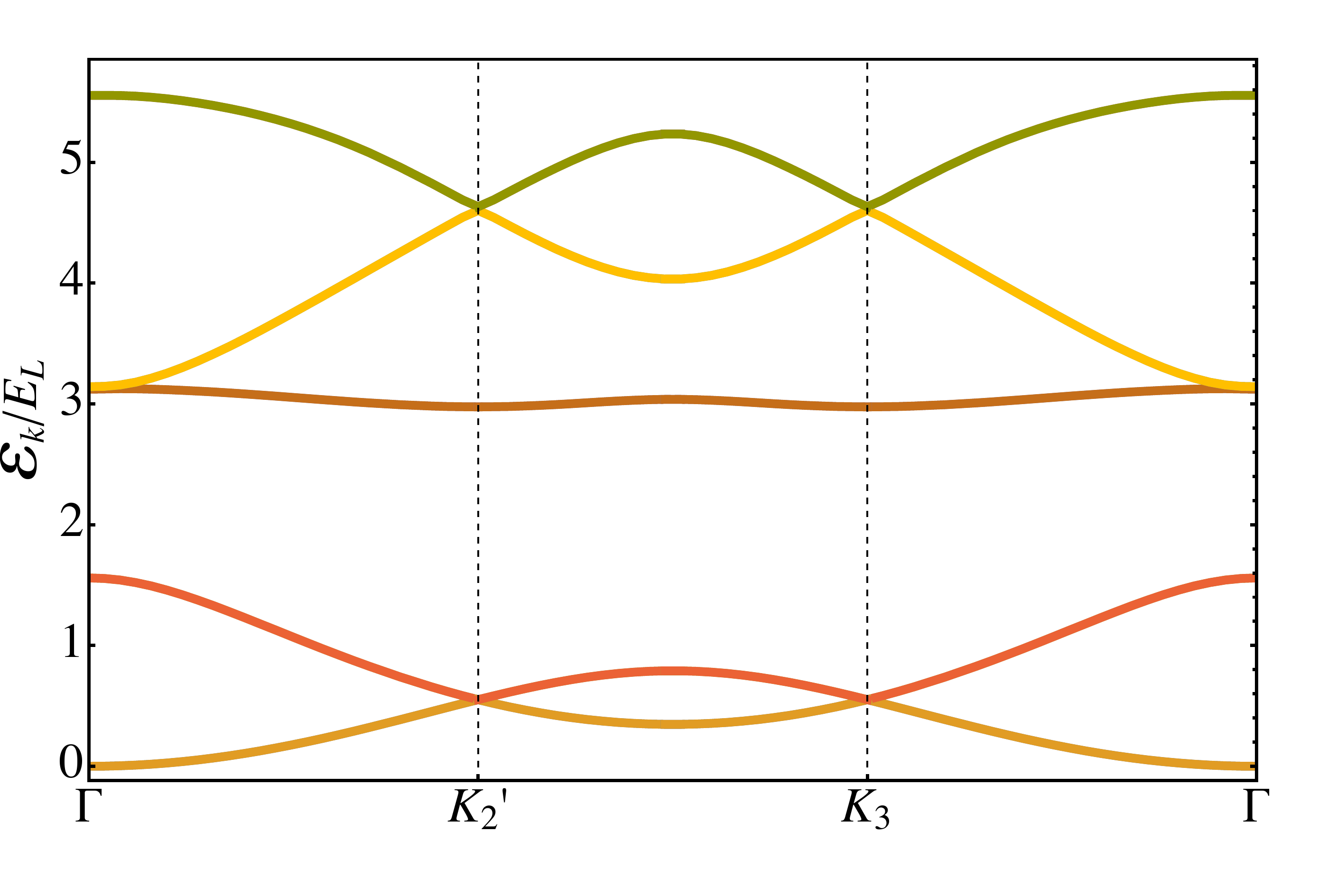}}
 {\includegraphics[width=0.3223\textwidth,clip]{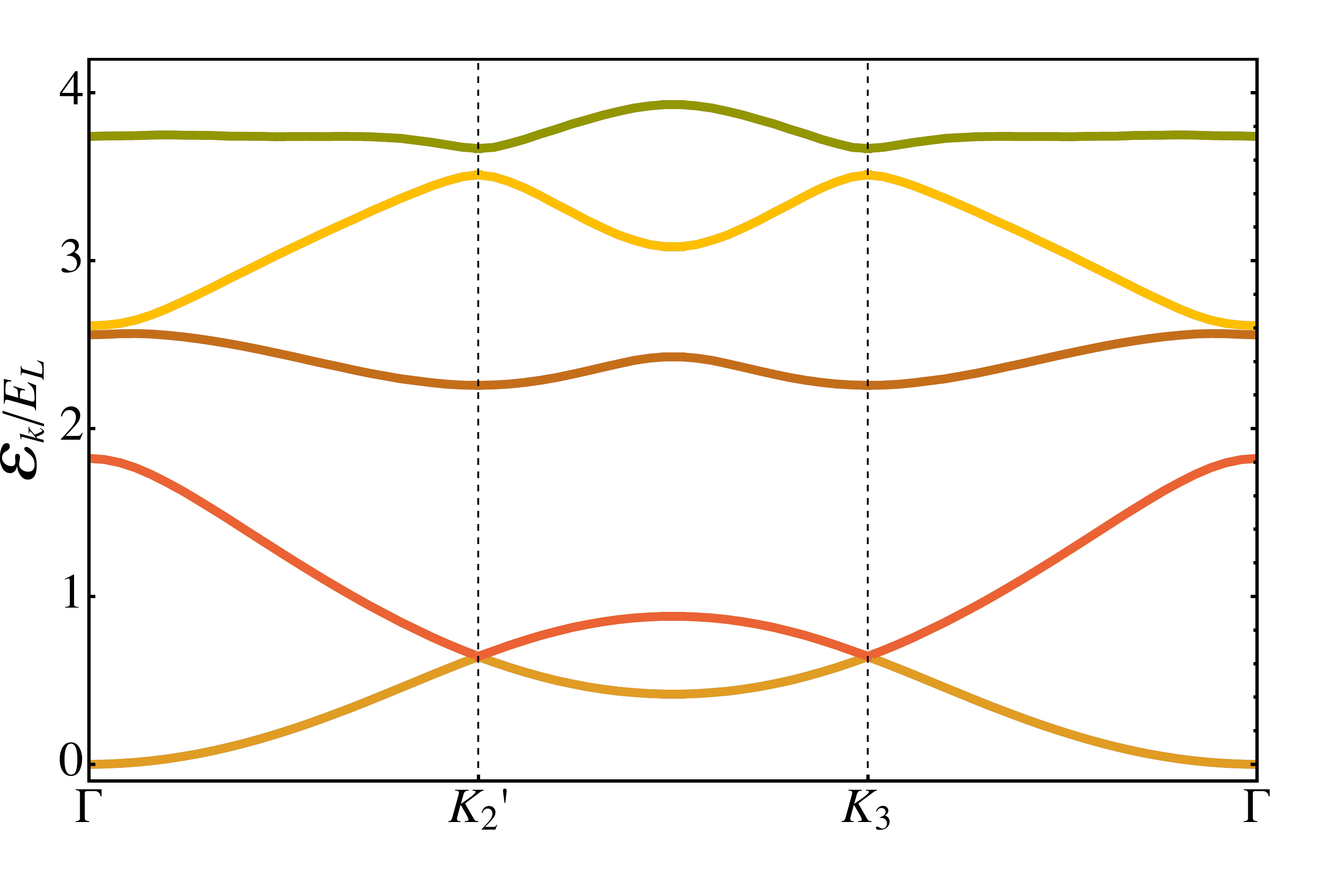}}
 {\includegraphics[width=0.3223\textwidth,clip]{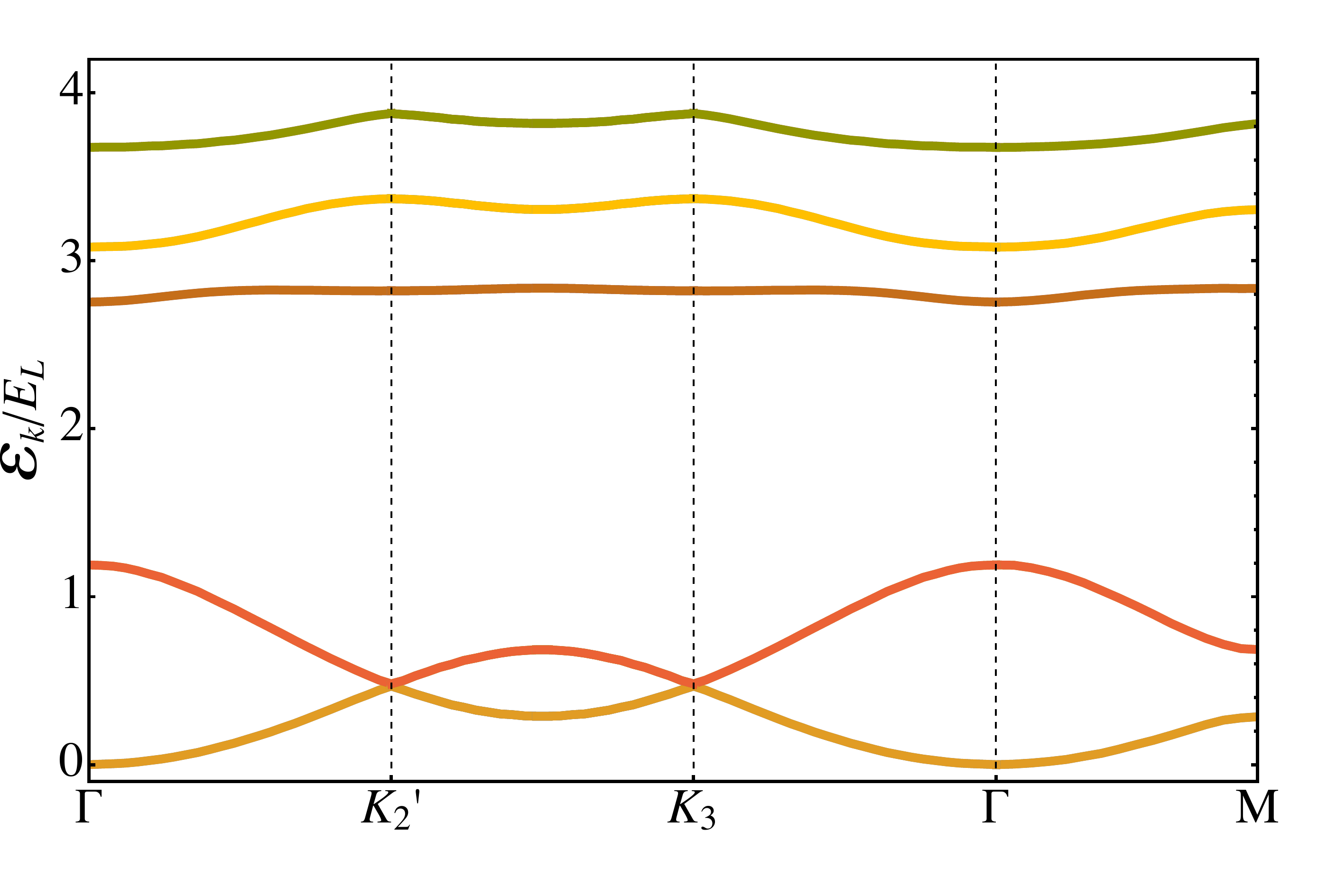}}
 \caption{Evolution of the mini-band spectrum upon tuning $d/L$ and $W$. All bands are doubly degenerate due to ${\cal T}$ and ${\cal P}$ symmetry; each band possesses a pair of Kramers spins, which we can refer to as spin $\uparrow$ and $\downarrow$. (a) Corresponds to the {\it Dirac regime}, with small spin-orbit interaction set by the small ratio $d/L=1/8$. We call the lower (upper) two mini-bands in the vicinity of the  Brillouin zone corners ${\bm K},{\bm K'}$, the first (second) Dirac points. The superlattice potential (\ref{U}) has strength $W=2E_L$, with $E_L$ from (\ref{EL}), chosen to produce steep second Dirac cones. (b) The parameters, $d/L=1.75/8$ and $W=E_L$, are chosen to produce a prominent TI gap at the second Dirac cones. (c) Demonstration of a nearly flat, topologically nontrivial band; the second highest in energy mini-band is seen to become nearly flat upon choosing parameters $d/L=1.75/8$ and $W=3E_L$, while the Chern numbers for each of the Kramers spins in this band are $C_{\uparrow,\downarrow}=\pm3$. Note: the topological band gaps at the first Dirac points are nonzero, yet smaller than the thickness of the lines, see Figure \ref{1DPGaps}(a).}
\begin{picture}(0,0) 
\put(-247,118){\text{(a)}} 
\put(-81,118){\text{(b)}}
\put(88,118){\text{(c)}}
\end{picture}
\label{BandEvolution}
\end{figure*}

\subsubsection{Nonperturbative, multi-band theory}
In the {\it multi-band approximation}, we use the exact diagonalization results for the Luttinger Hamiltonian (\ref{HL}) wavefunctions, which are compactly expressed in Eq. (\ref{ket}), and project the superlattice Hamiltonian operator (\ref{Hoperator}) onto this basis. Such a procedure generates the following matrix structure,
\begin{widetext}
\begin{align}
\label{multiband}
\notag {\cal H}(\bm k)_{(i,j),(l,m),(\sigma_l,\sigma_m)}&=\Bra{\bm k, j, l, \sigma_l} \hat {\cal H} \Ket{\bm k, i, m, \sigma_m}=\Bra{\bm k, j, l, \sigma_l} \hat {\cal E} \Ket{\bm k, i, m, \sigma_m} + \Bra{\bm k, j, l, \sigma_l} \hat{{\cal U}}({\bm r}) \Ket{\bm k, i, m, \sigma_m}\\
&={\cal E}_{i,l,\sigma_l}(\bm k_i)\delta_{i,j}\otimes\delta_{l,m}\otimes\delta_{\sigma_l,\sigma_m}+{\cal U}_{(i,j),(l,m),(\sigma_l,\sigma_m)}(\bm k_i)\delta_{\sigma_l,\sigma_m}\\
\notag {\cal U}(\bm k)_{(i,j),(l,m),(\sigma_l,\sigma_m)}&=W\sum_{t=1}^3\delta(\bm k_i-\bm k_j \pm \bm G_t)\sum_{S_z,S_z'}\sum_{n,n'}\hat{k}_{j,-}^{(\sigma_{l}-S_z')}\hat{k}_{i,+}^{(\sigma_m-S_z)} a_{l,n',S_z'}^*(|\bm k_j|)a_{m,n,S_z}(|\bm k_i|)\braket{S'_z,n'|S_z,n}\\
\notag&=W\sum_{t=1}^3\delta(\bm k_i-\bm k_j \pm \bm G_t)\sum_{S_z}\sum_{n}\hat{k}_{j,-}^{(\sigma_{l}-S_z)}\hat{k}_{i,+}^{(\sigma_m-S_z)} a_{m,n,S_z}^*(|\bm k_j|)a_{l,n,S_z}(|\bm k_i|)
\end{align}
\end{widetext}

\noindent where a given matrix element of the kinetic matrix ${\cal E}_{i,l,\sigma_l}(\bm k_i)$ is evaluated directly from the underlying 2DHG spectrum, also shown in Figure \ref{2DHGspectrum}(a). We note that previous approaches \cite{SushkovNeto2013} have taken a quadratic approximation, such that ${\cal E}_{i,l,\sigma_l}({\bm k}_i)={\bm k}_i^2/(2m^*)$, which we sketch via the dashed line in Figure \ref{2DHGspectrum}(a). To generate a large topological gap, we are required to consider momentum and energy scales that are beyond the validity of the single-band, quadratic approximation. Moreover, the wavefunctions obtained in perturbation theory only account for quadrupole mixing, which is only valid for small $k/k_0<1$, as seen in Figure \ref{2DHGspectrum}(b). This motivates our present construction over previous approaches \cite{SushkovNeto2013}.

That completes the mathematical preliminaries. We now proceed to our findings, which are arranged as follows: Section \ref{PT} considers the case of the ${\cal P}$ \& ${\cal T}$-symmetric topological insulator, looking at both the first and second Dirac bands, their edge modes and/or Chern numbers. Section \ref{P Broken} considers explicit ${\cal P}$-symmetry breaking and the influence on edge modes. This analysis is limited to the first Dirac band only. Section \ref{Conclusions} contains our conclusions and further discussion. Wherever possible, we present results in physical scales coincident with those currently experimentally achievable.

\section{Results: ${\cal P}$ \& ${\cal T}$-symmetric TI}\label{PT}

Our discussion will be centred around the so called first and second set of Dirac bands, which correspond to the lower and upper two bands, respectively, shown in Figure \ref{BandEvolution}. In particular, in the vicinity of the Brillouin zone corners ${\bm K},{\bm K'}$, it is also convenient to call the (nearly) band touching points of the first (second) Dirac bands the {\it first (second) Dirac points} -- or just 1DP (2DP). Moreover, our results are obtained by varying the two available tuning handles (recall from Section \ref{params}): the ratio $d/L$ and the strength, $W$, of the superlattice potential (\ref{U}). We will now discuss the influence of each. 



Tuning the ratio $d/L$ determines the energy/momentum scale at which the Dirac points occur relative to the underlying 2DHG spectrum, i.e. which part of the 2DHG spectrum is band folded at the Dirac point. As the ratio $d/L$ is increased from zero, the anti-crossing {\it kink} in the 2DHG spectrum, Figure \ref{2DHGspectrum}(a), moves from higher energies down towards the Dirac points of the corresponding band structure i.e. after imposition of the superlattice. At the same time, one can see that the wavefunctions are becoming maximally mixed at the scale $k\sim k_0$, Figure \ref{2DHGspectrum}(b), i.e. the pure spin projections $\pm3/2,\pm1/2$ are heavily mixed for $k\sim k_0$, due to spin-orbit coupling. Hence the location of the kink relative to the Dirac point provides a qualitative indication of the effective strength of the spin-orbit coupling at the Dirac point; the closer the kink the larger the spin-orbit coupling.  


The second tuning handle we have at our disposal is the parameter $W$. Tuning the energy scale of the potential $W$ relative to the characteristic energy of the Brillouin zone $E_L$ (\ref{EL}), provides a means to control the {\it steepness} of the Dirac cones, i.e the effective velocity in the vicinity of the Dirac points. For example, in Figure \ref{BandEvolution}(a) we choose $W=2E_L$ to approximately optimise the steepness of the 2DP -- this choice was also made in \cite{Tkachenko2015}. 

With the freedom of two tuning handles: $d/L$ and $W$, there are many quantitatively distinct band structures we can present. For conceptual clarity as well as for ease of presentation we discuss what we consider to be the two most important qualitatively distinct regimes: (i) The anti-crossing is band-folded to be in the vicinity of the first Dirac points (1DP), such that $k_0=2/d\approx K_{1DP}=4\pi/(3L)$, 
 and hence $d/L\approx1/2$. (ii) The anti-crossing is coincident with the second Dirac points (2DP), such that $k_0\approx K_{2DP}=2K_{1DP}$, and hence $d/L\approx1/4$. 


\subsection{Second Dirac Bands}
The current experimental limitations are approximately: $10\lesssim d \lesssim 30$nm and $L\gtrsim40$nm. Moreover, the energy scale of the problem is set by $E_0\sim1/d^2$ (or equivalently $E_L\sim1/L^2$ for fixed $d/L$), inspiring us to consider the lower limit of the well confinement $d\sim10$nm to maximise the topological band gaps.

Let us consider $L/d\sim4$, which is both (i) achievable experimentally, and (ii) places the anti-crossing at the 2DP. Figure \ref{BandEvolution} shows the evolution of the band structure with parameters $d/L=1/8, 1.75/8, 1.75/8$ and $W=2E_L,E_L, 3E_L$ for Figure \ref{BandEvolution}(a), (b) and (c), respectively. Figure \ref{BandEvolution}(a) corresponds the Dirac regime with vanishing spin-orbit gap; this result approximately coincides with the results obatined previously \cite{Tkachenko2015} assuming no spin-orbit interaction. Figure \ref{BandEvolution}(b) shows the opening of a significant topological gap at the 2DP, while the topological gap remains vanishingly small at 1DP (although it is still nonzero). This result has not been discussed previously, and represents one of our primary conclusions. From here we conclude that by tuning the Fermi energy to lie within the spin-orbit band gap of the the second set of Dirac bands represents a more suitable topological insulator than tuning to the first set of Dirac bands. {\it Suitability} here refers to the size of the topological gap and hence the states resilience to thermal fluctuations and disorder. The suitability is further supported by experimental density of holes; allowing for the Fermi energy to sit at the second Dirac bands accommodates higher densities and the current limit is $n\approx 10^{11}$ cm$^{-2}$, which already sits beyond the first set of Dirac bands. 

We find that the spin-orbit band gap $\Delta_{SO}$ (evaluated at the ${\bm K}$ points) of the 2DP is largest in Figure \ref{BandEvolution}(c), which corresponds to parameters $\{d/L,W\}=\{1/4,3E_L\}$. The gap is seen to be $\Delta_{SO}\approx0.5E_L$, which corresponds to $\Delta_{SO}\approx1.1$meV at $d=10$nm and $L=40$nm. Again, for fixed ratio $d/L$ the energy scale of the system $E_L\propto1/L^2$ and hence motivation for small quantum well confinement length $d$ and superlattice spacing $L$ is apparent. 

Each Kramers spin in the lower band of the second Dirac bands has Chern number $C_{\uparrow, \downarrow}=\pm3$, implying there are three pairs of topologically protected edge modes (when the Fermi energy is tuned to lie in the spin-orbit band gap of the second Dirac bands). The details of the Chern number calculation will be provided in section \ref{Edge}. Moreover, in Figure \ref{BandEvolution}(c) we see that the topologically nontrivial bands of the second Dirac bands are nearly flat. The nearly flat band generates a high density of states, and since the kinetic energy scale (band width) is vanishing the particle-particle interactions become important. Explicit calculation of particle-particle interactions or strong correlation effects is beyond the scope of the present work. However, on general grounds there is expected to be an instability towards an ordered strongly correlated phase with leading candidates; fractional TI \cite{Stern2009, NeuportPRB2011, Sun2011, Sheng2011, Zhong2013},  exotic ferromagnetism \cite{Katsura2010, Santos2012, Kumar2014, Doretto2015}, exotic charge density wave \cite{Kourtis2012, Hohenadler2013, Daghofer2014}. Here by exotic we mean non-trivial algebra \cite{Goerbig2012, PARAMESWARAN2013, Doretto2015} due to the topology of the band. To find which instability dominates, one needs to perform intensive numerical calculations. Alternatively, progress can be made assuming one such ground state and finding characteristic properties. We leave this programme for a future study. Ideally the present work will motivate experimental searches for the strongly-correlated physics.

\subsection{First Dirac Bands}
\begin{figure}[t]
{\includegraphics[width=0.234\textwidth,clip]{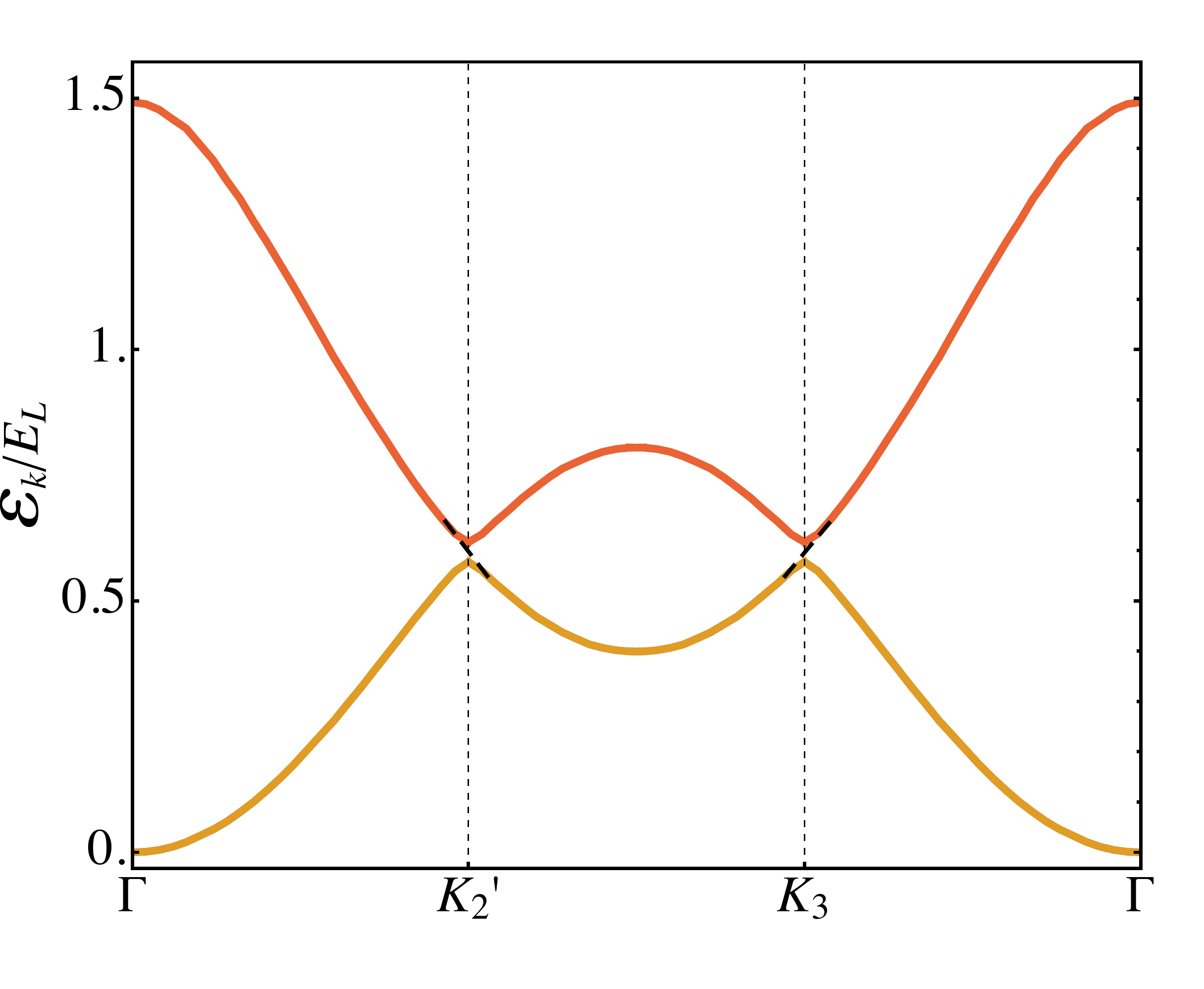}}
{\includegraphics[width=0.242\textwidth,clip]{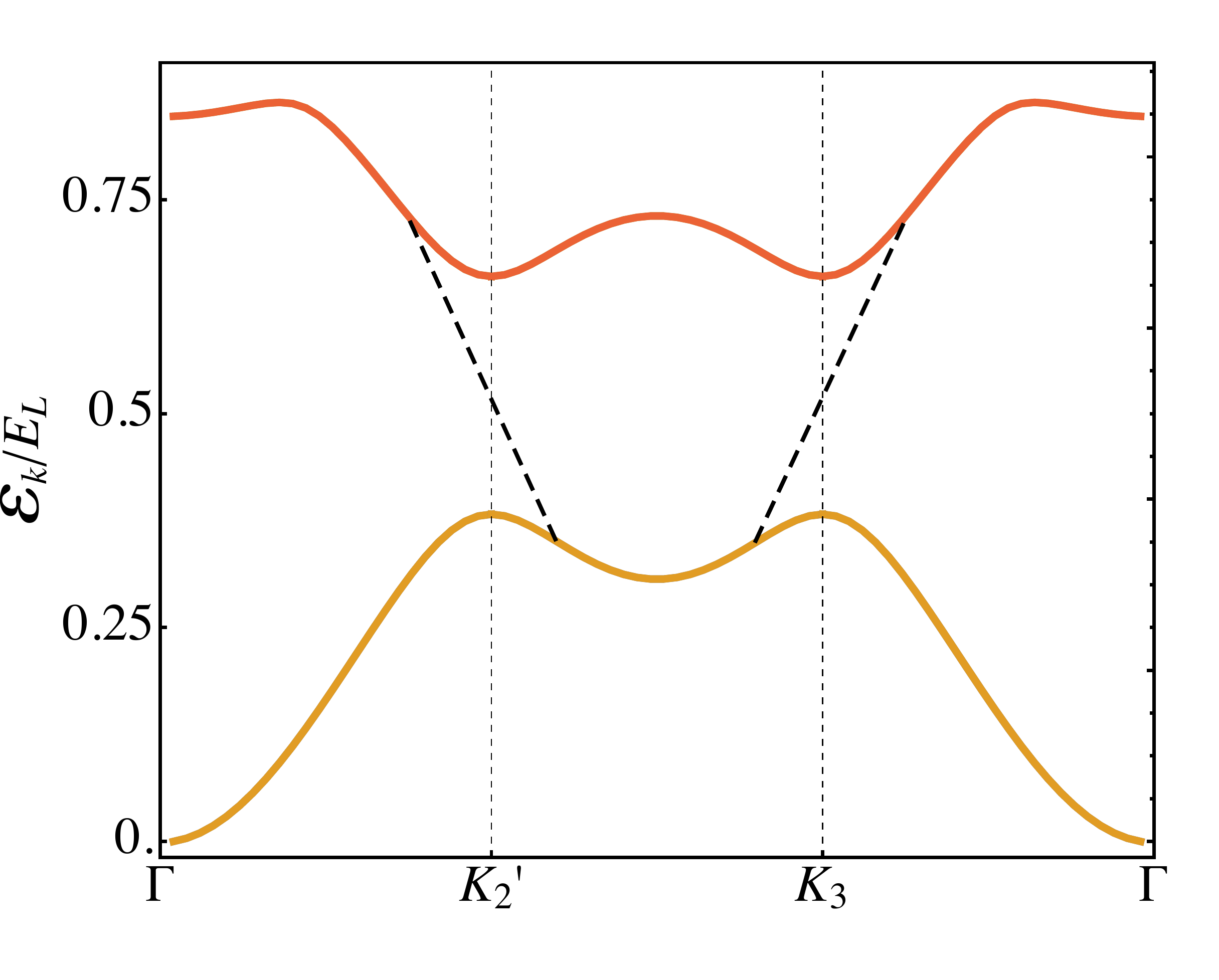}}
\caption{Comparison of the first Dirac points with parameters: (a) $d/L=1/4$,  (b) $d/L=1/2$, with the strength of the potential the same for both cases, $W=2E_{L=4}$.  The black dotted lines connecting the upper and lower bands in each figure are a schematic representation of the edge modes.}
 \begin{picture}(0,0) 
\put(-120,74){\text{(a)}}
\put(5,74){\text{(b)}}
\end{picture}
\label{1DPGaps}
\end{figure}
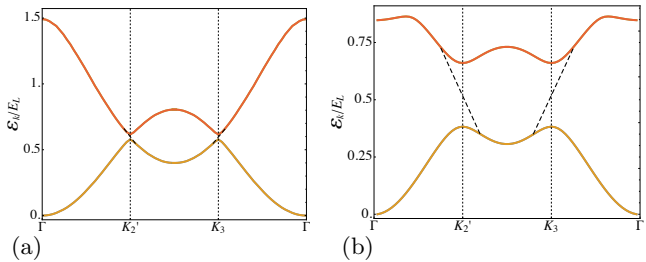

We now consider $L/d\sim2$, such that the anti-crossing kink of the underlying 2DHG dispersion is coincident with the first set of Dirac points. Consider Figure \ref{1DPGaps}(a), the topological gap at 1DP for $d/L=1/4$ are small -- as expected from our discussion above relating relative location of the anti-crossing kink and the Dirac point. Going to the regime whereby the kink corresponds to the 1DP, i.e. $L/d\sim2$ as shown in Figure \ref{1DPGaps}(b), the spin-orbit gap becomes large $\Delta_{SO}\approx0.3E_L$, which corresponds to $\Delta_{SO}\approx 0.8$meV if we choose feasible system parameters $d=20$nm and $L=40$nm. 

We comment on convergence of our multi-band approach, Eqs. (\ref{ket}) and (\ref{multiband}): for $d/L\lesssim1/4$ convergence is immediate between a one-band and two-band approach, with bands taken from the 2DHG dispersion Figure \ref{2DHGspectrum}(a). Going to larger ratios, say $d/L\sim1/2$, we find that we must include the third band of the 2DHG spectrum to reach adequate convergence. Note, it is possible to generate larger spin-orbit gaps by consider yet larger values of $d/L$ and $W$, but in such limits one must include more than three bands into the Hamiltonian (\ref{multiband}) to reach convergence. We limit ourselves to considering three bands, and so do not pursue this straightforward extension. 

\subsection{Edge Modes}\label{Edge}
To obtain the topological winding number of each band, we calculate the Chern number per Kramers spin species, i.e. one can define the Berry curvatures ${\cal F}_{x,y}^{b,\sigma_b}$, where $b$ labels the mini-band and $\sigma_b=\uparrow,\downarrow$ labels the two Kramers spins, directly from the Hamiltonian (\ref{multiband}) (or from its wavefunctions). As per usual, the Chern number for a given band and spin ($b,\sigma_b$) is then the integral of the Berry curvature over the entire Brillouin zone $C_{b,\sigma_b}=\int {\cal F}_{x,y}^{b,\sigma_b} dk_x dk_y/(2\pi)$. However, since we evaluate using a discrete $(k_x,k_y)$ momentum grid, it is convenient to employ the {\it lattice gauge theory} technique of Ref. \cite{Suzuki2005} to evaluate the $C_{b,\sigma_b}$. 

Due to ${\cal T}$-symmetry of the Hamiltonian ${\cal H}$ (\ref{multiband}), it follows that: (i) ${\cal H}$ can be made block diagonal in the Kramers spin index $\sigma_b$, and (ii) for a given band $b$, the Chern number is opposite in sign for each spin species, $C_{b,\downarrow}=-C_{b,\uparrow}$. The Chern numbers for the lowest four bands ($b\in\{1,4\}$) in Figure \ref{BandEvolution}(b)\&(c) are: $C_{b,\uparrow,\downarrow}=\mp1,\pm1,\pm0,\pm3$. 

Aside from the topological index, the edge modes have been calculated analytically for the 1DP in the perturbative approach of Ref. \cite{SushkovNeto2013}; there they find a single pair of counter propagating, opposite (Kramers) spin edge modes (as depicted schematically by the black dashed lines in Figure \ref{1DPGaps}), which is also consistent with our Chern number calculation $C_{\uparrow,\downarrow}=\mp1$ for the lowest band. 

Next we will perform an analogous semi-analytic calculation in the case of triangular well confinement (which generates a Rashba spin-orbit interaction). What one finds from the calculation to follow (or from \cite{SushkovNeto2013} in the absence of Rashba) is that the edge modes do not cross at the $\Gamma$ point (unlike in the Kane-Mele model of graphene \cite{Kane2005}). They are still ${\cal T}$-reversal symmetric partners, they just do not cross in momentum space. Without further calculation we suggest that this offers two advantages over usual graphene: First, backscattering from a ${\cal T}$-breaking impurity (i.e. magnetic impurity), must satisfy a strict momentum conservation condition and hence it is conceivable that such back scattering events have a restricted phase space, i.e. the topological edge modes are equipped with an extra {\it protection}. Second, in a finite geometry the overlap of the wavefunctions of edge modes which occupy opposite edges of the sample is expected to produce a finite gap in the edge mode dispersions due to level repulsion. By the same argument as for the case of backscattering, the non-crossing of the edge mode dispersions in momentum space reduces the possibility of the finite geometry-induced level repulsion.  We do not pursue these directions any further.


\section{Results: ${\cal P}$ Broken TI}\label{P Broken}
We now turn to the influence of explicit parity breaking. Employing the triangular well confinement (\ref{triconfine}), ${\cal T}$-symmetry remains intact while ${\cal P}$ is explicitly broken. Since almost all experimentally produced confining potentials posses some degree of inhomogeneity and with it ${\cal P}$-breaking (as well as that triangular wells are purposefully designed), we wish to understand how this ${\cal P}$-symmetry breaking affects the topological edge modes. 

We use a semi-analytic approach to elucidate the key influence of ${\cal P}$-breaking and Rashba spin-orbit coupling. As outlined in Ref. \cite{SushkovNeto2013} and section \ref{PertDirac}, we construct a low energy effective Hamiltonian in the vicinity of the ${\bm K}_j$ points (${\bm K'}_j$ is easily obtained thereafter), and is valid only for small ${\bm k}$ about this point, the result is shown in Eq. (\ref{Dirac}), see also the original work \cite{SushkovNeto2013}. Next, the ${\cal P}$-breaking (cubic) Rashba interaction is introduced via,
\begin{align}
\label{HR}
\delta H_R&=-\frac{i}{2}\alpha\left(k_+^3\tau_-  - k_-^3\tau_+\right)
\end{align}
where the raising operators $\tau_\pm=\tau_x\pm i\tau_y$ act on the two spins in the lowest $l=0$ subspace of the underlying 2DHG dispersion (i.e. the lowest dispersion branch of Figure \ref{2DHGspectrum}(a)), and $\alpha$ is an effective interaction constant. Evaluating the projection of $\delta H_R$ onto a plane wave basis, and following up with a projection into the pseudo-spin space (exactly following the steps described after Eq. (\ref{Dirac}) and given in more detail in \cite{SushkovNeto2013}), we obtain the effective Hamiltonian,
\begin{align}
\label{HR1}
H_R&=-v(k_x\sigma_z + k_y \sigma_x)\otimes{\mathbb I} - \eta\sigma_y\otimes\tau_z - \gamma\mathbb I\otimes\tau_y.
\end{align}
Here $\gamma\sim \alpha 8K^3$ (with $K=4\pi/(3L)$) is due to the Rashba spin-orbit term (\ref{HR}) and is an energy scale that is comparable to the Rashba splitting $\Delta_R$ shown in Figure \ref{Triangularspectrum}(b). Again, this Hamiltonian is valid for ${\bm k}\sim0$, and is an expansion about the ${\bm K}$ points; taking $v\to-v$ and $\gamma\to-\gamma$ one obtains the corresponding expansion about the ${\bm K'}$ points. We easily obtain the four eigenvalues of the effective Hamiltonian (\ref{HR1}),
\begin{align}
{\cal E}_k&=\pm\sqrt{(v {\bm k}\pm\gamma)^2+\eta^2}.
\end{align}
Hence the (gapped) Dirac cones are shifted from the ${\bm K}$-points ($\bm k=0$) to $v {\bm k}\pm\gamma=0$. It is easy to deduce the corresponding eigenvalues at the $K'$ point, just by demanding the ${\cal T}$-symmetry condition: ${\cal E}^\uparrow_k={\cal E}^\downarrow_{-k}$. 

\begin{figure}[t]
  {\includegraphics[width=0.235\textwidth,clip]{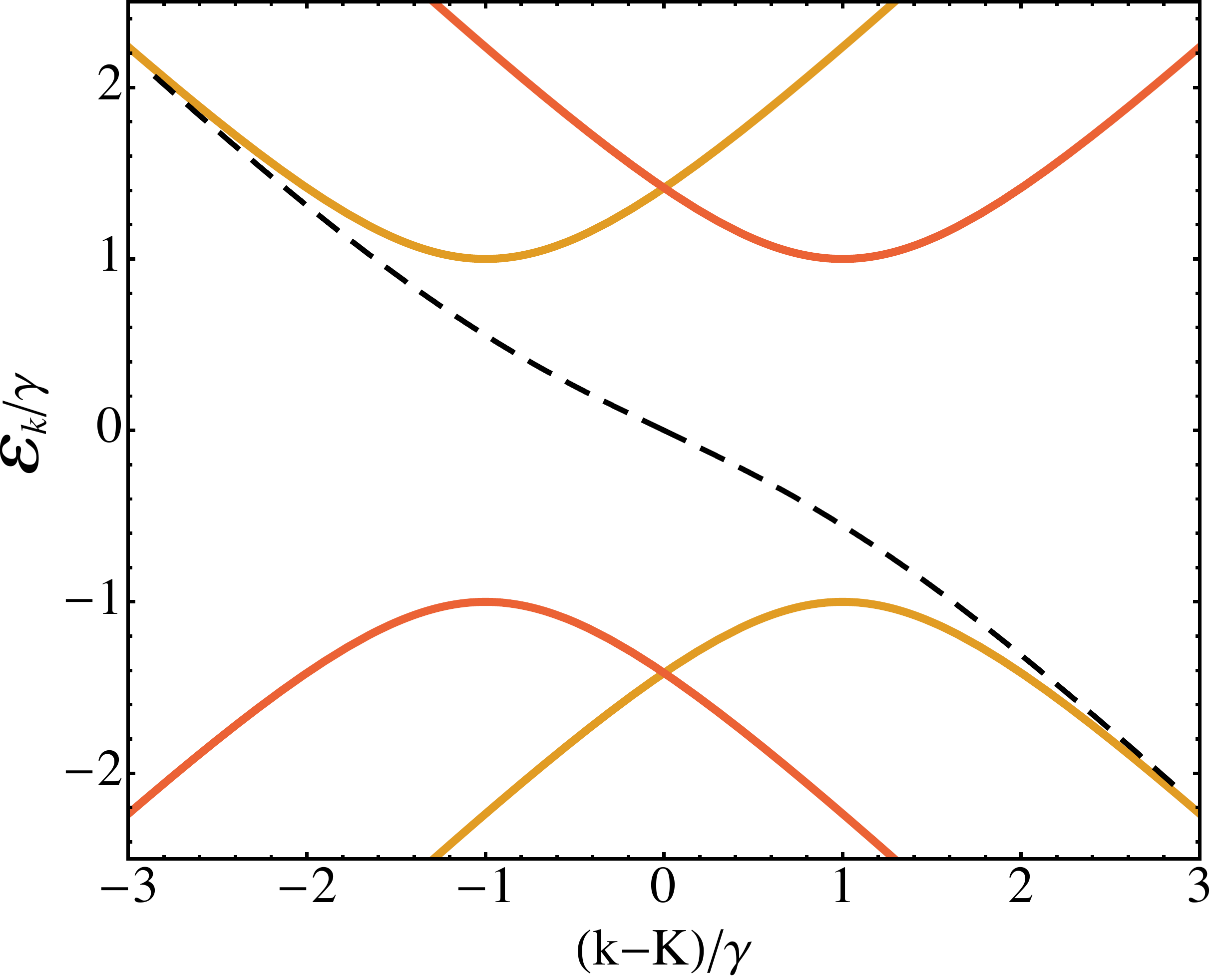}}
    {\includegraphics[width=0.235\textwidth,clip]{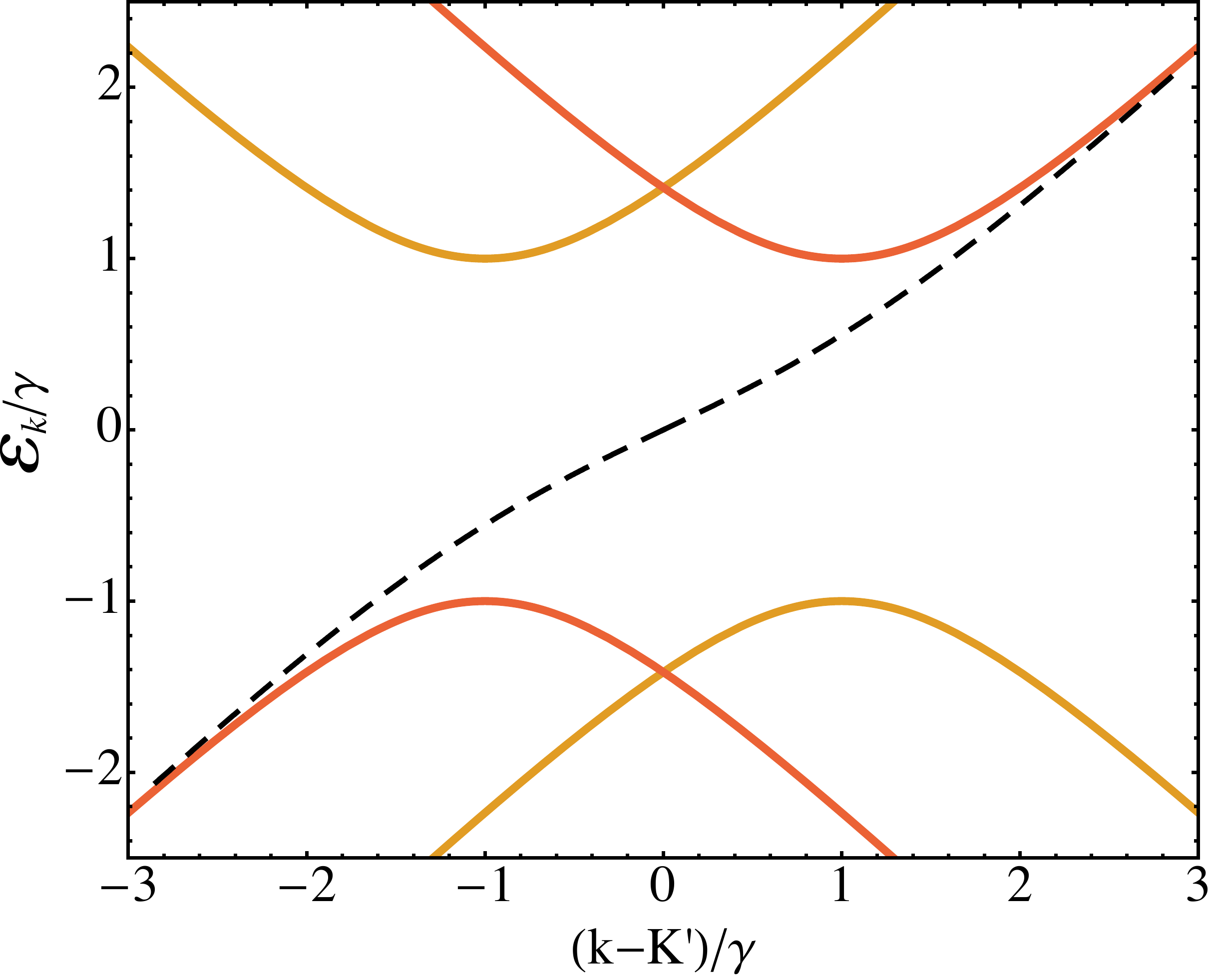}}
 \caption{Effect of Rashba (${\cal P}$-breaking) on the band structure. The bulk (solid lines) and edge (dashed lines) bands are evaluated in the vicinity of (a) the ${\bm K}$ points, and (b) the ${\bm K'}$ points. We take the parameters $\eta=\gamma=v=1$ in Eq. (\ref{HR1}).}
 \begin{picture}(0,0) 
\put(-117,64){\text{(a)}}
\put(7,64){\text{(b)}}
\end{picture}
\label{RashbaEdgeModes} 
\end{figure}
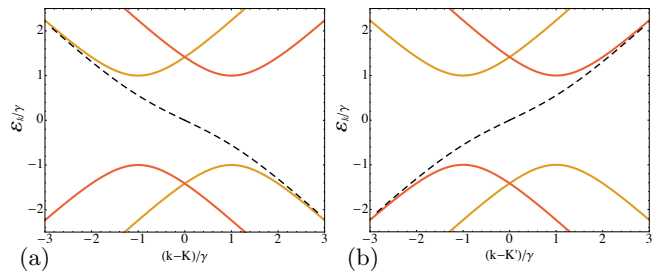

To find the dispersions of edge modes in the low energy Hamiltonian description (\ref{HR1}), we follow the techniques of \cite{SushkovNeto2013, Li2016} and impose a hard wall boundary condition. We obtain the edge mode dispersions numerically, and so we do not present an equation here, instead the results for the particular set of parameters $\eta=\gamma=v=1$ in Eq. (\ref{HR1}) are shown in Figure \ref{RashbaEdgeModes}. The low energy Hamiltonian (\ref{HR1}) is only reliable for small ${\bm k}$ expansions about either ${\bm K}$ or ${\bm K'}$, and Figure \ref{RashbaEdgeModes}(a) and (b) show separately the expansion about ${\bm K}$ and ${\bm K'}$. 
We refer the reader, interested in the details of the edge mode calculation, to the appendix of Ref. \cite{Li2016}, which contains the generalization of \cite{SushkovNeto2013} sufficient to handle the present case. 

Overall, qualitatively, we find that the two degenerate copies of band spectrum under ${\cal P}$-symmetry, are now non-degenerate, and are simply momentum-shifted copies of each other. Most importantly, the topologically protected edge modes remain in tact. Of course, taking into account momentum dependence of the Rashba term (\ref{HR}), see also Figure \ref{Triangularspectrum}(b), through the entire BZ will quantitatively change this conclusion from being a uniform momentum space shift, to a momentum dependent shift. But the salient point remains: breaking ${\cal P}$-symmetry does not destroy the TI state constructed here. We therefore need not worry about experimental inhomogeneities, or the influence of non-rectangular quantum well confinement geometries. 

\section{Conclusions}\label{Conclusions}

We consider the topological insulating states of {\it artificial graphene} -- generated by imposing a superlattice structure on top of a two dimensional hole gas in a semiconductor heterostructure -- 
and develop a method to calculate the band structure. The method developed and presented here provides a non-perturbative treatment of the spin-orbit interaction and, in particular, is not limited to low energy/momentum scales. Previous approaches had precisely this limitation \cite{SushkovNeto2013, Li2016}. 

Using the developed technique we discuss the previously found {\it first set of Dirac points} (1DP), and confirm that they indeed represent a topological insulating state. Moreover, we point out an extra {\it protection} of the edge modes against magnetic impurities and finite geometry effects. We also perform a semi-analytic, perturbative calculation to elucidate the effect of explicit parity (inversion) symmetry breaking, taking the particular case of a triangular well confinement. The analysis shows that parity breaking does not destroy the key properties of the topological edge states. 

Our most important findings relate to what we call the {\it second set of Dirac points/bands}; these Dirac points sit at twice larger energy/momentum scales than the first set, and so have been completely beyond the validity of previous approaches \cite{SushkovNeto2013, Li2016}. Our developed technique is indeed appropriate to describe the second Dirac bands, and we find the following desirable properties: (i) Owing to sitting at higher energy/momentum, the second Dirac bands experience larger spin-orbit interaction than the corresponding first Dirac bands, and hence posses a larger topological/spin-orbit band gap. With the present experimental limitations on the length scales $d$ and $L$, this makes the topological insulating state obtained by tuning the Fermi energy to the second Dirac bands more robust than the topological insulating state obtained by tuning the Fermi energy to the first. 
(ii) The second Dirac bands posses three pairs of counter propagating edge modes -- Chern number per spin species is $C_{\uparrow,\downarrow}=\pm3$. Compare with the first Dirac bands, which have just a single pair of counter propagating edge modes ($C_{\uparrow,\downarrow}=\mp1$). (iii) Upon tuning the system parameters, we demonstrate the appearance of nearly flat bands endowed with a nontrivial topology ($C_{\uparrow,\downarrow}=\pm3$). This finding suggests that hole-hole interactions become the dominant energy scale and as a result the system is expected to exhibit strongly correlated phases -- most notably, a fractionalised topological insulator state, superconductivity, or exotic forms of magnetism.  

In conclusion, the present work has exploited the highly tuneable nature of artificial graphene to show that it is an excellent candidate to realise (i) a topological band insulator phase, and (ii) a plethora of enigmatic, strongly correlated phases.

We thank Dima Miserev for his invaluable insights into the Luttinger Hamiltonian, and Raja Grewal for general discussions. This research was fully supported by the Australian Research Council Centre of Excellence in Future Low-Energy Electronics Technologies (project number CE170100039) and funded by the Australian Government.


\end{document}